\documentclass[aps,prb,groupedaddress,twocolumn]{revtex4}
\usepackage{color}
\usepackage{graphicx}
\usepackage{dcolumn}
\usepackage{bm}
\usepackage{ulem}

\begin{document}

\title{Local SiC photoluminescence evidence of non-mutualistic hot spot formation and sub-THz coherent emission from a rectangular Bi$_2$Sr$_2$CaCu$_2$O$_{8+\delta}$ mesa}

\author{Hidetoshi Minami$^{1-3}$}
\email[]{minami@bk.tsukuba.ac.jp}
\author{Chiharu Watanabe$^{1,3}$}
\author{Kota Sato}
\altaffiliation{Present Address; Institute for Solid State Physics, University of Tokyo, Kashiwa, Chiba 277-8581, Japan}
\author{Shunsuke Sekimoto$^{1,3}$}
\author{Takashi Yamamoto}
\altaffiliation{Present Address; Wide Bandgap Materials Group, Optical and Electronic Materials Unit, Environment and Energy Materials Division,
National Institute for Materials Science, 1-1 Namiki, Tsukuba, Ibaraki, 305-0044, Japan}
\author{Takanari Kashiwagi$^{1-3}$}
\author{Richard A. Klemm$^4$}
\author{Kazuo Kadowaki$^{1-3}$}

\affiliation{$^1$Graduate School of Pure \& Applied Sciences, University of Tsukuba, Tennodai, Tsukuba Ibaraki 305-8573, Japan\\
$^2$Division of Materials Science, Faculty of Pure \& Applied Sciences, University of Tsukuba, Tennodai, Tsukuba, Ibaraki 305-8573, Japan\\
$^3$CREST-JST (Japan Science \& Technology Agency)\\
$^4$Department of Physics, University of Central Florida,  Orlando, Florida 32816-2385 USA}

\date{\today}


\begin{abstract}
>From the photoluminescence of SiC microcrystals uniformly covering a rectangular mesa of  the high transition temperature $T_c$ superconductor Bi$_2$Sr$_2$CaCu$_2$O$_{8+\delta}$, the local surface temperature $T({\bm r})$  was directly measured  during simultaneous sub-THz emission from the $N\sim10^3$ intrinsic Josephson junctions (IJJs) in the mesa. At high bias currents $I$ and low bath temperatures $T_{\rm bath}\lesssim~35$ K, the center of a large elliptical hot spot with $T({\bm r})> T_c$  jumps dramatically with little current-voltage characteristic changes.  The hot spot doesn't alter the ubiquitous primary and secondary emission conditions: the ac Josephson relation and the electromagnetic cavity resonance excitation, respectively.  Since the most intense sub-THz emission was observed for high $T_{\rm bath}\gtrsim~50$ K  in the low $I$ bias regime where hot spots are absent,  hot spots cannot provide the primary mechanisms for increasing the output power, the tunability, or for promoting the synchronization of the $N$ IJJs for the sub-THz emission, but can at best coexist non-mutualistically with the emission.  No $T({\bm r})$ standing waves were observed.
\end{abstract}

\maketitle
\section{Introduction}
Application of a dc current $I$ to a superconductor can result in extreme spatial variations in the local temperature $T({\bm r})$ known as hot spots\cite{SBT,Huebener,Eichele,Dieckmann,Gross,Haugen,Niratisairak1,Niratisairak2,Wang1,Wang2,Guenon,Benseman13}.  A hot spot was first inferred from the observed hysteresis in the current-voltage $I$-$V$ characteristics (IVCs) of low transition temperature $T_c$ superconducting microbridges\cite{SBT}.  Subsequently thin film hot-spot $T({\bm r})$'s were observed using low-$T$ scanning electron microscopy (LTSEM)\cite{Huebener, Eichele}.  In devices of the high-$T_c$ superconductor YBa$_2$Cu$_3$O$_{7-\delta}$ (YBCO), equivalent hot spot detections were made using LTSEM and low-$T$ scanning laser microscopy (LTSLM)\cite{Dieckmann}.  Both LTSEM and LTSLM scanning measurements can distort the $T({\bm r})$ observations because they  do not measure $T(\bm{r})$ directly but measure overall voltage changes $\delta V({\bm r})$  induced by the local heating of either the perturbing scanning electron or laser beams\cite{Eichele}, and their interactions with $T({\bm r})$.  Secondly, from their experimental principles, $T(\bm{r})$ obtained by LTSEM and LTSLM can be exaggerated near the transition temperature $T_c$ because of the peaking response at ${T_c}$\cite{Huebener,Eichele,Wang1,Gross}.  Moreover, they are also surface sensitive because of the different absorption coefficients depending on the different surface materials.  Especially, the LTSLM images may include spurious reflections due to the excessive responses from the edges of the mesa.\cite{Gross,Wang1,Wang2,Guenon}.

 Kolodner and Tyson invented another method to measure the local temperature with high spatial resolution\cite{Kolodner-Tyson1,Kolodner-Tyson2}.  It   involves covering the sample with a thin polymer film containing rare earth complexes, exciting the complexes with ultraviolet (UV) light, and measuring their resulting $T$-dependent photoluminescence (PL).   By applying this technique to thin films of YBCO and the high-$T_c$ superconductor Bi$_2$Sr$_2$CaCu$_2$O$_{8+\delta}$ (Bi2212), the photoluminescence of the covering polymer film provided detailed maps of the hot spot $T({\bm r})$ in the superconducting film without any scanning distortions, because this technique allows for a direct measurement of $T({\bm r})$\cite{Haugen,Niratisairak1,Niratisairak2,Benseman13}.  However, even such measurements can include $T(\bm{r})$ distortions caused by multiple light reflections from parallel regions of the polymer film surface and variable fluorescence efficiencies arising from regions of non-uniform polymer film thickness, especially at the sample's corners and edges\cite{Benseman13}.

 These undesired distortions possibly present in the measured $T(\bm{r})$  can be eliminated by a new PL technique in which the sample surface is uniformly covered with a large number of SiC microcrystals\cite{Choyke,Konstantinov,DevatyChoyke,Fan}.  The irregular microcrystals have  sizes of $2~\sim~3~\mu$m, comparable to the sample thickness, so surface reflections  are randomly oriented and edge distortions are negligible.  After exposure to UV light, the SiC microcrystals emit strongly $T$-dependent PL\cite{Choyke,DevatyChoyke,Fan}, which may be used to calibrate the $T(\bm{r})$ measurements.  SiC forms in 3C, 4H, 6H, 15R, and less common polytypes, with energy gaps of $\sim$ 2 eV\cite{DevatyChoyke,Fan}.  The challenges are to find the polytype mixture with a strongly and continuously $T$-dependent PL over the $T$ range from $T\ll T_c$ to $T\gg T_c$, to produce a uniform coverage of the superconductor with SiC microcrystals, and to find the best circumstances for their use.

Here we employ SiC microcrystals to study the relationship between the simultaneous formation of hot spots and the coherent sub-THz emission from Bi2212 mesas.  The recent discovery of  coherent sub-THz electromagnetic (EM) wave emission from  mesa structures of Bi2212 has generated  great interest\cite{Ozyuzer, Kadowaki}.  The goal is to fill the so-called ``THz gap''\cite{Tonouchi} with the intrinsic Josephson junctions (IJJs) stacked one-dimensionally along the $c$ axis of a single crystal\cite{Ozyuzer, Kadowaki, Bulaevskii, Lin}. Since there are $N\sim 670$ identical IJJs in a single crystalline Bi2212 mesa 1 $\mu$m thick, the coherent output power, $P$ $\propto N^2$ from a mesa can be enhanced from $\sim~$pW for a single junction to tens of $\mu$W\cite{Ozyuzer,Kadowaki,Turkoglu,Yamaki,Sekimoto,Benseman,Yamamoto}.  However,  Bi2212 mesa structures with $N>10^3$ IJJs encounter serious heating problems due to the large $I$ injected into the mesa\cite{Ozyuzer}, causing the mesa to develop a hot spot $T({\bm r})$ that exceeds $T_c$ on its top surface, even though the bath temperature $T_{\rm bath}$ of its substrate is held to well below $T_c$\cite{Gross,Wang1,Wang2,Guenon,Benseman13}.

This hot-spot behavior  and its effects upon the sub-THz emission from Bi2212 were studied by LTSLM\cite{Gross,Wang1,Wang2,Guenon}. Based upon their observations that the sub-THz emission can occur even with such a highly nonuniform $T({\bm r})$ variation containing a local hot spot with $T({\bm r})>T_c$ in the higher $I$ bias regime of the IVCs, they suggested that the hot-spot formation may be beneficial to the synchronization of the $N$ IJJs, an essential ingredient for obtaining intense sub-THz radiation.  Numerical simulations of a hot spot\cite{Yurgens}, the fact that the $c$-axis resistance $R_c(T)$ can be shunted by the hot spot due to $T$ regions of negative differential resistivity, $d\rho_c(T)/dT<0$\cite{Yurgens,Gross}, the much narrower line width of $\lesssim~23~$MHz in the high-$I$ bias hot spot region than in the retrapping, low-$I$ bias regime\cite{Li,Kashiwagi}, and the lack of radiation from mesas with thicker Au electrodes\cite{Kakeya}, all tend to support their hypothesis that the hot spots play an essential role in the synchronization of the IJJs.

Besides the hot-spot detection, the LTSLM measurements also detected standing waves in rectangular and disc-shaped mesas\cite{Wang1, Wang2, Guenon}.  In describing the observations, it was suggested that at high-$I$ bias, the mesa was effectively divided into  a hot-spot region where $T({\bm r})>T_c$ and the superconducting remainder of the mesa\cite{Wang1, Wang2, Guenon}.  It was further suggested that sub-THz radiation arising in the remaining superconducting region at a frequency $f=f_J=(2eV)/(hN)$ due to the ac Josephson effect, where $e$ is the electronic charge and $h$ is Planck's constant, could be enhanced by locking onto an effective EM cavity mode frequency $f_{\rm cav, eff}\approx c_0/(2n\ell_{\rm eff})$\cite{cavitymode},  where $c_0$ is the speed of light in vacuum, $n~\sim~4.2$ is the index of refraction of Bi2212, and the  standing wavelength $2\ell_{\rm eff}$ is determined  by the effective EM cavity geometry generated by the boundaries of the single superconducting region of the mesa reduced in size by the $T({\bm r}_b)=T_c$ boundaries of one or more hot spots\cite{Wang2,Guenon}.  If this suggestion were correct, then changing  the  locus  of the hot-spot boundaries could increase the tunability of the mesa's sub-THz emission $f$.

In rectangular mesas, the standing waves observed either by LTSLM or by rare-earth embedded polymer films formed along the mesa's length\cite{Wang1,Wang2,Guenon,Benseman13}, although usually the observed $f=f_{\rm cav}=c_0/(2nw)$\cite{cavitymode}, by excitation of the TM(1,0) thin rectangular EM cavity mode, where $w$ is the mesa width both in the high-$I$ bias reversible (R-type) region and in the low-$I$ bias retrapping irreversible (IR-type) regions, regardless of hot spot formation.\cite{Ozyuzer,Kadowaki,Yamaki,Sekimoto,Yamamoto,Benseman,Kashiwagi,Minami2}.  Unfortunately, angular distribution experiments in the planes normal to the length and width of the rectangular mesa could not distinguish the difference between those associated standing wave models\cite{Klemm}.  In one disk mesa, a standing wave-like pattern similar to that of the TM(3,1) thin cylindrical EM cavity mode was detected by LTSLM\cite{Guenon}, which they claimed to be standing magnetic field waves\cite{Guenon},  not standing electric field waves that could be generated by the observed $\delta V({\bm r})$ changes\cite{Eichele}. But it was not clear whether the observed emission was generated by the excitation of that mode, by the previously detected fundamental TM(1,1) mode\cite{Tsujimoto1}, or by something else.   A model calculation with  an inhomogeneous critical current  simulating a hot spot   did not provide support to those observations\cite{Asai}.  By application of a magnetic field $\mu_0 H=0.0058$ T parallel to the layers, the standing waves disappeared, and were therefore interpreted as arising from the Josephson effect\cite{Wang1}. Thus the questions of whether the standing waves detected by LTSLM or (by the PL measurement with rare-earth embedded polymer film\cite{Benseman13}) were EM cavity waves associated with the emission, and/or were only a spurious EM wave effect generated by the laser itself (and/or its surface reflections), and/or were also $T({\bm r})$ standing waves, or were yet something else, were unresolved.

Here we present direct SiC microcrystalline PL measurements of $T({\bm r})$ on the top of a rectangular Bi2212 mesa during sub-THz emission, that settles the main question of whether the hot spot plays an important role in synchronizing the IJJs for the sub-THz emission, and the secondary question of whether the  standing waves observed by LTSLM or rare-earth embedded polymer films were EM cavity waves associated with the emission, were just a spurious EM wave effect, were $T({\bm r})$ standing waves\cite{Wang1,Wang2,Guenon,Benseman13}, or were yet something else.  Microcrystalline SiC PL experiments directly detect $T({\bm r})$ on the $\mu$m length scale but do not detect EM standing waves. In agreement with the LTSLM and rare-earth embedded polymer thin film experiments\cite{Wang1,Wang2,Guenon,Benseman13}, the SiC PL measurements revealed that a rectangular mesa can have an extremely non-uniform $T({\bm r})$ distribution forming an elliptical hot spot with a central $T({\bm r})>T_c$ region, while the rest of the mesa is still superconducting.

The size of the hot-spot region in the mesa grows or shrinks reversibly in the higher-$I$ bias regime of the IVCs, and sometimes the hot-spot position jumps dramatically from one central location ${\bm r}_0$ to another within the mesa without significant IVC changes, and it suddenly vanishes and $T({\bm r})$ becomes almost uniform when $I$ is reduced to its lower bias regime.  We further observed strong sub-THz radiation at rather high-$T_{\rm bath}$  $\gtrsim$ 50 K values  in the retrapping regions of lower-$I$ bias, where hot spots do not exist, suggesting strongly that hot spots do not play a crucial role in the synchronization of $N$ IJJs for the sub-THz radiation.  Moreover, in striking contrast to the standing waves observed by LTSLM and rare-earth embedded polymer films\cite{Wang1,Wang2,Guenon,Benseman13}, $T({\bm r})$ standing waves were never observed in our SiC PL experiments.

The sub-THz radiation frequency $f$ was observed to  be essentially unaffected by the size and position of the hot spot, as it always maintains the primary and secondary radiation conditions: the  ac-Josephson relation $f=f_J=(2eV)/(hN)$ and the rectangular mesa EM cavity TM(1,0) resonance condition $f=f_{\rm cav}=c_0/(2nw)$, regardless of a hot spot's presence\cite{Ozyuzer,Kadowaki,Yamaki, Sekimoto,Benseman,Yamamoto, Kashiwagi,Minami2}.  Considering all of the experimental results described above, whether this non-mutualistic relationship between the hot spot and the sub-THz emission extends to the remarkable synchronization of the emissions from the $N$ IJJs is still unknown\cite{Kashiwagi,Li}, but our results strongly suggest that a mutualistic synchronization effect is unlikely.\cite{Comment}

\section{sample preparation and experimental methods}
\subsection{Mesa preparation and microcrystalline SiC coating}
The sub-THz emitting mesa devices were fabricated by an Ar-ion milling technique using metal masks from a piece of a mm-sized thin single crystal of Bi2212\cite{Mochiku}.  More details were given previously\cite{Kashiwagi,Minami1,Minami2}.   Of the several candidate $T({\bm r})$-sensitive coating materials we tested, the air-stable, commercial (Kojundo Chemical Laboratory Co. Ltd., Japan) mixed-polytype SiC powder 2 - 3 $\mu$m in size was chosen because it exhibited not only the strongest $T$ dependence of its PL intensity near the $T_c$ of Bi2212, but also the best non-magnetic, chemically resistant and electrically insulating properties.  A water suspension of the SiC microcrystals was painted on the top surface of the mesa, and the sample was turned upside down to dry.  While drying, gravity and the water surface tension combined to spread the microcrystals uniformly on the hanging suspension surface.  After drying, the uniformity of the microcrystalline SiC surface coverage was checked by direct observation of the PL intensity under a microscope by exposing UV light over the entire mesa at $I=0$, as described in detail in Section IIC.

Figure 1(a) was produced from  a photograph of the device studied, which consists of five congruent rectangular mesas of length $L$ = 400 $\mu$m and width $w$ = 80 $\mu$m.  One mesa was measured by atomic force microscopy to have a trapezoidal cross section with top and  bottom widths of 79 and  89 $\mu$m, respectively, and a height of $d$ = 2.40 $\mu$m, including a 70 nm thick Au top electrode. The small red square highlights the current injection position. Figure 1(b) is a stereographical sketch of the mesa device highlighted by the large red rectangle in Fig. 1(a).

\begin{figure}[t]
\includegraphics[width=8.6cm, trim=0.5cm 0.5cm 1cm 0.5cm, keepaspectratio, clip]{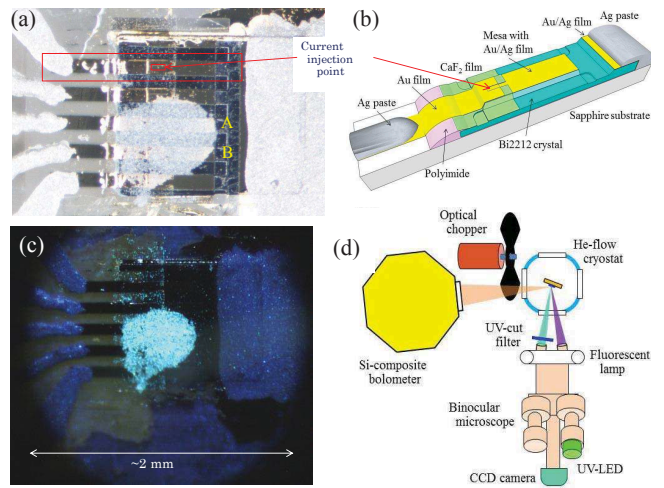}
\caption{\label{fig1}(color online) (a) A photograph of the  device containing five rectangular Bi2212 mesas fabricated  on  a  single crystal.  The SiC microcrystals form the lower white blob.  The central mesa A is covered rather uniformly, whereas the lower right part of mesa B (just below mesa A) is  uncovered.    The small red rectangle  indicates the $I$ injection point.  (b) A stereographical drawing of the mesa indicated by the large red rectangle in Fig. 1(a).  (c) A photograph of the sample under UV light with $\lambda_{\mathrm{center}}\sim$ 375 nm at $T_{\rm bath}$ = 15 K.  (d) A sketch of the optical setup for simultaneous PL and sub-THz radiation measurements.}
\end{figure}

\subsection{Experimental setup and the radiation power}
UV light with $\lambda_{\mathrm{center}}\sim$ 375 nm obtained from a commercial light emitting diode (LED) was uniformly cast upon the device  within a circular area of radius  $\sim$ 1 mm centered on the SiC coated area.  In Fig. 1(c), the dark violet areas arise from the diffuse reflection of Ag paste.  The central bright light-blue area arises from the SiC  PL.

In Fig. 1(d) the  optical setup is sketched.  The UV-LED was mounted in a view port of the binocular microscope, and the UV light passed through an objective lens and the quartz cryostat window in order to irradiate the sample inside the He-flow cryostat.  The PL was led to the charge-coupled device (CCD) camera by passing through the same cryostat quartz window, a UV-cut filter and another objective lens.  After passing through a polyethylene cryostat window and a previously described optical chopper\cite{Kadowaki}, the sub-THz  mesa emission was simultaneously detected by a Si-composite bolometer  and by a Fourier transform infrared (FTIR) spectrometer (JASCO, FARIS-1)\cite{JASCO}, as described in the following and elsewhere.\cite{Watanabe}

The radiation power of the mesa used in this experiment was estimated as follows.  Due to the location of the photoluminescence imaging equipment, the incident angle from the sample mesa to the Si-bolometer was obliged to be very shallow, about 70$^{\circ}$ from top of the mesa.  At this detection angle, the signal intensity is strong enough for detection, but is reduced by an order of magnitude from that optimally obtainable. The abscissa in Fig. 3(c) and the ordinates in Fig. 3(b) and Fig. 5(b) present the direct output values of the lock-in amplifier.

The total radiation power was calculated from its angular dependence as  was done  previously\cite{Kashiwagi}.   The responsivity of the Si-bolometer used in the present experiments is $\sim$11.0 mV/nW, the solid angle at the distance of 8 cm from the sample to the window of the Si-bolometer is 0.005 sr, the preamplifier gain is 200, the transmission efficiency of the chopper is about 50 \%, and accounting for the attenuation of the three quartz windows, the total power was estimated to be $\sim$5 $\mu$W.  This value is  an order of magnitude lower than the maximum power observed so far.  We noticed that this is due partly to the effect of the SiC coating\cite{Watanabe}.

We define  $\hat{\bm x}$, $\hat{\bm y}$, and $\hat{\bm z}$ to be unit vectors along the mesa's width, length, and height, as sketched in Fig. 1 of Ref. [36].  Then the angle $\theta$ varies from $+90^{\circ}$ when the emission is  along $\hat{\bm x}$, to $-90^{\circ}$ when the emission is along  $-\hat{\bm x}$.  Using these coordinates, UV light was irradiated at an incident angle of  $\theta\sim~25^{\circ}$ from $\hat{\bm z}$ towards $\hat{\bm x}$, and the PL and the sub-THz radiation were observed at the angles of $\theta\sim~15^{\circ}$ from $\hat{\bm z}$ towards $\hat{\bm x}$, and $\theta\sim~-70^{\circ}$, or $70^{\circ}$ from $\hat{\bm z}$ towards  $-\hat{\bm x}$, respectively.

\begin{figure}[t]
\includegraphics[width=8.6cm, trim=1cm 4cm 1.5cm 1cm, keepaspectratio, clip]{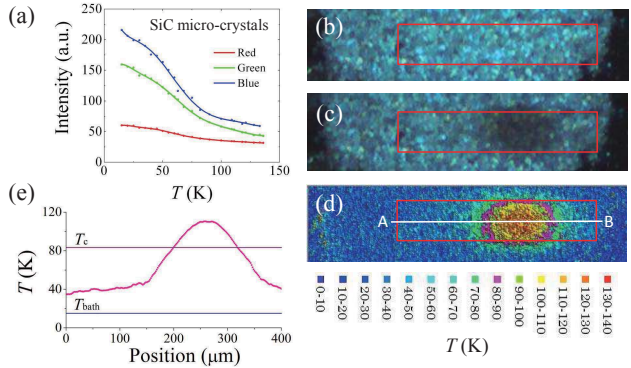}
\caption{\label{fig2}(color online) (a) The $T$ dependence of the red, green, and blue  components  of the observed PL intensity integrated over the mesa surface with $I=0$.   (b) and (c) The expanded PL images of mesa A at $T_{\rm bath}$=15 K  at $I=0$ and at $I$ = 40.1 mA, respectively.  (d) The analyzed $T({\bm r})$ map of the mesa with $T_{\rm bath}$=15 K, $I$=40.1 mA, and $V$=1.53 V. The red rectangles in (b)-(d) outline the mesa. (e) A $T(x)$ profile of mesa A along the AB line shown in Fig. 2(d).  The solid curve is obtained by numerical averaging.}
\end{figure}

\subsection{SiC PL intensity calibration and conversion of SiC PL image to $\bm{T}(\bm{r})$}
 In order to test the uniformity of the SiC microcrystalline coverage of mesa A, it was subjected to UV light at $T_{\rm bath}=15$ K with $I=0$, and an image of the resulting PL is shown  in Fig. 2(b).   It was observed here that the PL intensity on a scale of 10 $\sim$ 20 $\mu$m is not uniform, although the coverage of the SiC microcrystals appears rather uniform as seen in Fig. 1(c).  Moreover,  detailed examination of Fig. 2(b) on the scale of a few $\mu$m reveals coarse bright regions corresponding to the PL from individual microcrystals.  Because of the non-uniform PL response and the spotty image due to the coarse microcrystalline grain size, a direct conversion of the observed local PL intensity to $T({\bm r})$ maps would result in significant scatter in the  $T({\bm r})$ map data on such a fine scale.   This artificial noise effect can be minimized as much as possible by dividing the PL image from the microcrystals covering the mesa surface area and its surroundings into 80 $\times$ 350
pixels and individually calibrating the PL intensity measured by each pixel  as a function of temperature as described below.

To calibrate the SiC PL intensities with $T$, their integrated  red, green and blue visible frequency components are separately measured for 15 K $\le T_{\rm bath}\le 138$ K with $I=0$ and displayed in Fig. 2(a).  Since  the blue component has the largest $T$ sensitivity, it was used for the measurements.  The actual PL images observed from the microcrystals covering mesa A and its surroundings at $T_{\rm bath}$ = 15 K with $I=0$ and   $I$ = 40.1 mA are shown in Figs. 2(b) and 2(c), respectively.  Because of the hot-spot formation due to the local heating, the PL intensity from that region is reduced, in accordance with the local $T({\bm r})$ rise, as indicated for $T({\bm r})$ with $I=0$ in Fig. 2(a), yielding the dark region of higher $T({\bm r})$ to the right side of the center of the red rectangle in Fig. 2(c).  Each of the 80$\times$350 pixels in the overall image contains the  detected PL intensity of each of the three primary colors with 8-bit resolution.  After the PL images were measured during simultaneous sub-THz radiation observations at various bias currents, we took the PL images at $I=0$ at every 4 $\sim 5^{\circ}$ K under the same UV irradiation conditions, and the blue components of the accumulated images were compared with those of the images taken under an $I$ bias at each pixel of the corresponding position one by one to determine the $T({\bm r})$ map of the biased mesa.  The spatial and relative $T(\bm{r})$ resolutions are limited to 2-3 $\mu$m and to a few K, respectively.


\begin{figure*}[t]
\includegraphics[width=15cm, trim=1cm 2.5cm 1cm 1cm, keepaspectratio, clip]{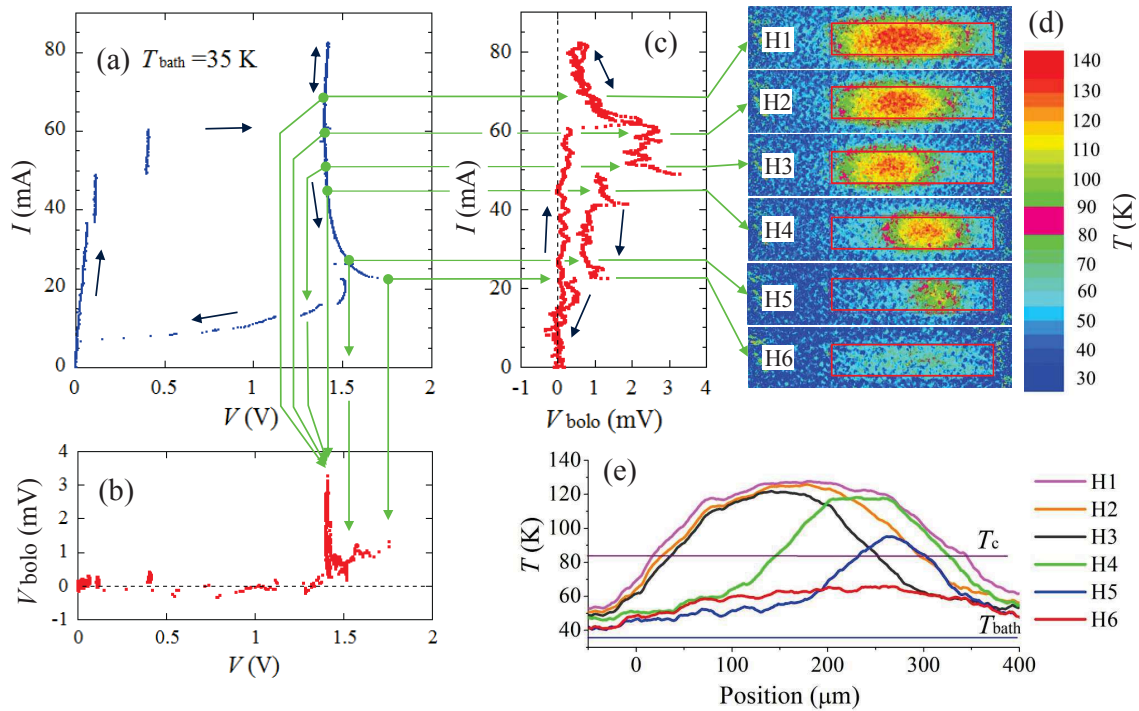}
\caption{\label{fig3}(color online) (a) The IVCs of mesa A at $T_{\rm bath}$ = 35 K.  (b) and (c) The radiation power (output voltage of the lock-in amplifier) detected by the Si-composite bolometer plotted versus $V$ (b) or $I$ (c).  (d) The $T({\bm r})$ maps observed at 6  $I$ levels (H1-H6) with  color coding.  (e) The numerically-averaged $T({\bm r})$ profiles of the hot spot along the length similar to Fig. 2(e) for the H1-H6 data.}
\end{figure*}

\section{experimental results}
\subsection{Observation of a local hot spot}
Figure 2(d) represents an example of a $T({\bm r})$ map of mesa A obtained with $I$ = 40.1 mA and $V$ = 1.53 V at  $T_{\rm bath}$ = 15 K, indicating a hot spot with $T({\bm r})>T_c$.  First of all,  the maximum $T({\bm r}_0)$ of about 120 K at the center ${\bm r}_0$ of the hot spot  is clearly observed, while both mesa edge regions, especially the larger left edge, have  $T({\bm r})\sim$~40 K, also exceeding $T_{\rm bath}=$ 15 K.  Such an extremely non-uniform $T({\bm r})$ agrees at least qualitatively with the former LTSLM and polymer film results\cite{Wang1,Wang2,Guenon,Benseman13} and even quantitatively with computer simulations\cite{Yurgens}, where the current density of 133 A/cm$^2$ in Fig. 2(a) of Ref. [29] corresponds to $I$ = 43 mA for the present mesa.  In contrast to the extremely non-uniform $T({\bm r})$ in the lateral directions, we note that the $T$ gradient along the $c$ axis can primarily be neglected within the mesa height of 2.40 $\mu$m according to  numerical simulations\cite{Yurgens}. This extremely inhomogeneous $T({\bm r})$ can be clearly visualized in Fig. 2(e), where  $T(x)$  along the  AB line in Fig. 2(d) is shown.  We note that the hot spot is not located at the $I$ injection point, indicated by the small red square in Fig. 1(a), but is shifted by about 10 \% of the distance from the center of mesa A to its right end, differing from the LTSLM results\cite{Wang1, Wang2}, but simulations indicated that this could be a geometrical effect\cite{Gross}.

Another surprise is that the sub-THz radiation is essentially unaffected by these drastic thermal inhomogeneities, except at IVC points of jumps or disappearances of a hot spot, as noted in Section IIIB.  In Fig. 3(a) the IVCs (blue curves) of mesa A at $T_{\rm bath}$ = 35 K biased with a constant $I$ mode are shown together with the simultaneous recordings of the Si bolometer output $V_{\rm bolo}$ versus $V$ and $I$,  as respectively shown in Figs. 3(b) and 3(c).  We note that the amplitude of the background thermal radiation is small, ($V_{\rm bolo}$ $\approx$ 0.5 mV at most) even at $I$ = 82 mA,  compared with the sub-THz radiation intensity of a few mV as seen in Fig. 3(c).  We recall that this radiation intensity $V_{\mathrm{bolo}}$ measured by the bolometer  is observed at $\theta$ = 70$^{\circ}$ from the $c$-axis due to the experimental constraints, which intensity is approximately an order of magnitude reduced from the maximum value that could be observed at some other angle.  The $T({\bm r})$ maps at 6 of the more than 20  mapped IVC points are presented as panels H1-H6 in Fig. 3(d).  Upon lowering $I$ from 82 mA, the measurable onset of  the sub-THz radiation appears at around $I$ = 70 mA corresponding to panel H1, where the hot spot with $T({\bm r})>T_c$ occupies more than 85 \% of the mesa surface.  The sub-THz radiation intensity then increases rapidly with decreasing $I$, as seen in Fig. 3(c).  The $T(x)$ profiles along the length similar to that in Fig. 2(e) are plotted in Fig. 3(e).

\subsection{Jumping hot spots}
The hot-spot area gradually decreases as $I$ is reduced.  In some cases minor changes in the IVCs cause the hot-spot center ${\bm r}_0$ to move dramatically, as if by jumping, as seen by comparing panels H3 and H4 in Figs. 3(d) and 3(e).  Although a discontinuous change in the sub-THz radiation intensity is associated with dramatic hot-spot center ${\bm r}_0$ jumps at $I$ = 49 mA, the remaining sub-THz emission properties such as the IVCs and $f$ are essentially unchanged by jumping of the hot spot position.  Such dramatic jumping hot spots are more frequently observed at $T_{\rm bath}$ = 25 K\cite{Watanabe}.   Here, we note that the hot spot has a tendency to appear near the mesa center, but the radiation intensity appears to increase when the hot spot moves  from the center toward a short edge (or end).  These features will be shown more clearly in a forthcoming publication.\cite{Watanabe}  Furthermore, as seen by comparing panels H5 and H6 in Fig. 3(d) as well as Fig. 3(e), the hot spot vanishes rather sharply and transforms to a nearly uniform $T({\bm r})$ state as $I$ is decreased below 22.4 mA.  It is interesting to mention that the $I$-$V$ curve has a distinct jump only when the hot spot appears or disappears, but it has only very little effect when the hot spot moves in the mesa.

The intentional manipulation of the hot spot position was reported previously by injecting current into the mesa from different edges or by adjusting the current to  multiple electrodes attached at different places to the mesa, which generated the hot spot at different positions\cite{Guenon}.   Although those authors observed the wavy features by controlling the position of the hot spot (for example, see Fig. 7 in reference 8) it is unfortunately difficult to compare our results with theirs without knowing the spectral information, \textit{i.e.}, the intensity, the frequency, the line width, \textit{etc}.

\begin{figure}[t]
\includegraphics[width=8.6cm, trim=1cm 0.5cm 1cm 0cm, keepaspectratio, clip]{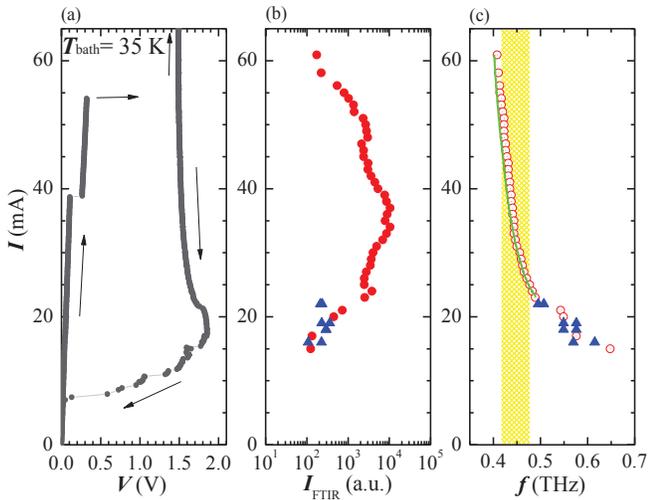}
\caption{(color online) (a): IVCs of mesa A at $T_{\rm bath} =$ 35 K.  (b): The spectral intensity (peak height) $I_{\mathrm{FTIR}}$ obtained is shown on a logarithmic scale when $I$ is decreased from 61 mA to 15 mA.  Below $I=22$ mA, $I_{\rm FTIR}$ splits into two peaks as represented by the blue triangles.  (c) The open red circles and also the filled blue triangles below $I=22$ mA, represent the radiation $f$ values directly measured by the FTIR.  The thin green curve indicates the calculated $f_J=2eV/(hN)$ for the  $V$ in Fig. 4(a) and the fixed number $N$ = 1.59$\times 10^3$ of IJJs.  The shaded yellow region highlights the calculated region of expected EM cavity resonance enhancements accounting for the trapezoidal mesa cross section.}
\label{fig4}
\end{figure}

\subsection{Spectral intensity measurements}
In Fig. 4, the voltage $V$, the spectral intensity $I_{\mathrm{FTIR}}$ measured  by an FTIR spectrometer, and $f$ are given for mesa A as functions of $I$ at $T_{\mathrm{bath}}$ = 35 K.  We note although the mesa and $T_{\rm bath}$ are the same as in Fig. 3, various physical parameters such as the IVCs, $I_{\mathrm{FTIR}}$ and $f$ differ slightly in each experimental run due to different heat conduction conditions of the mesa.  Because of this, the IVCs in Fig. 4(a) are slightly different from those of Fig. 3 even though both data were taken at  $T_{\rm bath} = 35$ K. In particular, we also note that the hot spot behavior (not shown here) in the run corresponding to Fig. 4  was also not the same as that shown in Fig. 3(d), \textit{i.e.}, the jumping phenomena of the hot spot observed in panels H3 and H4 of Fig. 3(d) were not observed in the run corresponding to Fig. 4\cite{Watanabe}.  When $I$ is significantly larger than 62 mA, Joule heating causes  $T({\bm r} )> T_c$ for the entire mesa, so that there is no sub-THz radiation except for that from thermal radiation associated with the normal state.  As detailed in Fig. 4 when $I$ is gradually decreased from higher values to $\sim$~61 mA,  a small portion of the mesa, especially including its short edges as shown in panel H1 of Fig. 3(d), becomes superconducting, and weak sub-THz radiation becomes observable.  As $I$ is further decreased from 61 mA to 52 mA, corresponding respectively to  panels H1 to H2 in Fig. 3(d),  $I_{\mathrm{FTIR}}$ grows by an order of magnitude.  Note that the ordinate in Fig. 4(b) is shown on a logarithmic scale, while it is on a linear scale in Fig. 3(b) and 3(c).  In this region of the IVCs, the hot-spot area with $T({\bm r})>T_c$ shrinks and the combined area of the  separate superconducting regions with $T({\bm r})<T_c$ expands.  However, even though the superconducting regions are separated by the central hot spot, the mesa radiation $f$ spectrum still exhibits a single peak with a width limited by the instrumental resolution of 7.5 GHz.  Upon further decreasing $I$ from 45 to 35 mA, $I_{\mathrm{FTIR}}$ increases by another order of magnitude and reaches a maximum for $I$~$\sim$~35 mA, roughly corresponding to panel H3 of Fig. 3(d).  Similar results with decreasing $I_{\mathrm{FTIR}}$, and decreasing area with $T({\bm r})>T_c$ are observed as $I$ is decreased from 35 to $\sim$ 23 mA, as indicated by panel H4 in Fig. 3(d) and curve H4 in Fig. 3(e), and in panel H5 in Fig. 3(d) and curve H5 in Fig. 3(e).  Upon further decreases in $I$ below 22 mA where the hot spot has disappeared, the $I_{\rm FTIR}$ spectrum splits into two weak peaks, as observed previously\cite{Li}, the amplitudes of which rapidly decrease with decreasing $I$, becoming indistinguishable from that of noise for $I \sim$ 14 mA.  In this region 22 mA $\gtrsim I \gtrsim ~$15 mA,  $T({\bm r})$ of the mesa decreases nearly uniformly, without any hot spot.  In contrast to the results shown in Fig 3(c), a sudden decrease of the THz emission intensity associated with jumping hot spot phenomena  was not observed in this run.  More details of the relations between the  hot spot and the emission $I_{\mathrm{FTIR}}$ and $f$, including the effects of hot spot jumping, will be shown elsewhere\cite{Watanabe}.   Overall, when $I$ is reduced from 61 mA to 15 mA, $I_{\mathrm{FTIR}}$ shows a broad maximum centered around 35 mA, where the maximum signal to noise ratio is $\sim$ 500, as shown in Fig. 4(b).

Since the R-type IVCs for the $I$ range 61 mA $ \gtrsim I \gtrsim $~23 mA are continuous and reversible  with increasing or decreasing $I$\cite{Yamaki,Minami2},  the total number $N$ of IJJs is considered to be invariant.  However, for $I \lesssim$ 22 mA, the IR-type IVCs are irreversible with increasing and decreasing $I$. This irreversibility is due to the retrapping phenomenon associated with the non-linear character of the active IJJs accompanied by a sudden decrease in their active number $N$.  At $T_{\rm bath} = 35$ K as in Fig. 4, this irreversibility starts with the sudden disappearance of the hot spot and the IR-type of sub-THz radiation is not seen except for $I \lesssim$ 22 mA.  In contrast, almost only IR-type radiation was observed from the same mesa at $T_{\rm bath} = 55$ K, as shown in Fig. 5.

\subsection{Obedience to the ac-Josephson relation}
We emphasize that in all of the cases we studied, including the R- and IR-types of radiation in the same mesa as described above, the radiation frequency $f$ always equals  $f_J =2eV/(hN)$. Since  $V$ changes with $I$ in the overall IVCs, $f$ also changes with $I$, when $I$ is varied.  Assuming $N$=1.59$\times 10^{3}$ is fixed over the R-type radiation region 62 mA $\gtrsim I \gtrsim$ 23 mA  at $T_{\rm bath} = 35$ K,   $f_J(I)$ as obtained from the IVCs in Fig. 4(a) is plotted  in Fig. 4(c)  as the thin green curve, along with the  $f(I)$ points observed by the FTIR spectrometer, which are shown by the open red circles.  The curve and the symbols agree very well in this R-type radiation region.

When $I$ is decreased from the maximum applied value of 61 mA,  at which the hot spot occupies most of the mesa at $T_{\rm bath} = 35$ K, $f$ increases from its lowest value of 0.415 THz   with  increasing $V$ and decreasing $I$, as seen in Fig. 4(c).   This tunable increase is continuous until $f=0.49$ THz.  At $I$ = 21 mA, $V$ jumps to a higher value as seen in Fig. 4(a), corresponding to $I$ = 22.4 mA in Fig. 3(a), where the hot spot vanishes rather sharply and the mesa's  $T_{\rm bath}<T({\bm r})<T_c$ becomes nearly uniform.  This results in an effective increase in $V/N$ and hence in $f_J$, and  normally causes a corresponding increase in $f$ as seen in Fig. 4(c).  As $I$ decreases from 20 mA to 15 mA, $f$ begins to increase rapidly, though $V$ only decreases slightly, resulting in a correspondingly stronger decrease in $N$.  Below $I$ = 15 mA as the retrapping phenomenon  occurs repeatedly, $V$ correspondingly jumps several times to a lower $V$ value   before reaching $V=0$.  Note that $f=f_J$ even in this retrapping region, as previously shown explicitly for the internal IVC branches of different mesas\cite{Tsujimoto2}.  Therefore, even in the IR-type radiation region of the IVCs,  $f=f_J$ is still fixed to $V$, but it is not directly related to $I$, as its only relation to $I$ is through the IVCs.  Hence, we conclude that $f=f_J$ is satisfied in all radiating IVC regions of every mesa, as expected theoretically.

What is remarkable is that as the thermal condition of the mesa varies dramatically as $I$ is varied from 70 mA to 15 mA while holding $T_{\rm bath}$ fixed at 35 K, the large hot spot occupies most of the area of the mesa over the higher-$I$ region between 70 mA and 55 mA, the area of the hot spot shrinks gradually over the intermediate range of the current between 55 mA and 23 mA, and  at the lower end of the current variation at 22 mA, the hot spot vanishes and the mesa attains a more homogeneous $T_c>T({\bm r})$.  Although such catastrophic changes in the mesa thermal conditions occur, $f=f_J$ as seen in Fig. 4(c).  It is even more surprising that the radiation intensity $I_{\mathrm{FTIR}}$ shows a maximum in the middle $I$ range near to 35 mA, where intermediate hot-spot $T({\bm r})>T_c$ areas, 20-40 \% area of the total area of the mesa surface, are clearly observed.  $T({\bm r})$ in that $I$ range varies and corresponds roughly to those depicted in panels H3 and H4 in Fig. 3(d).

\begin{figure*}[t]
\includegraphics[width=15cm, trim=1cm 2cm 2cm 0cm, keepaspectratio, clip]{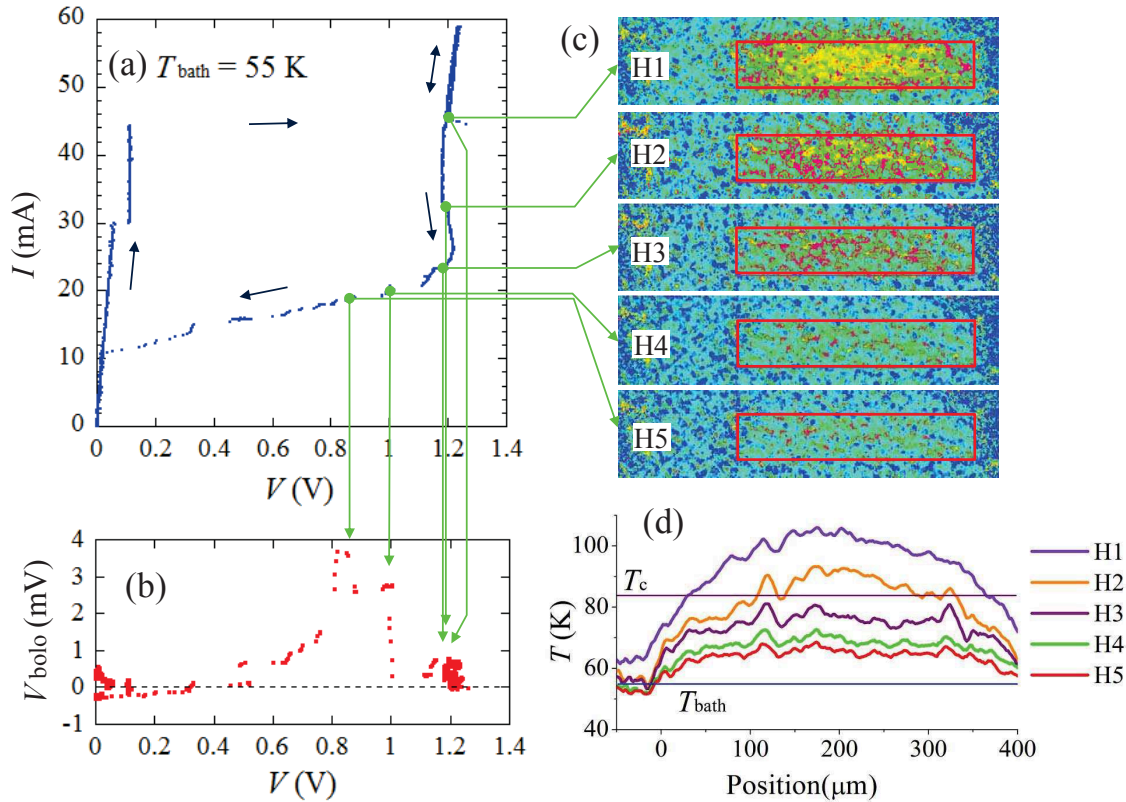}
\caption{\label{fig4}(color online) (a) The IVCs of mesa A at $T_{\rm bath}$ = 55 K.  (b) The radiation power versus $V$.  (c) The $T({\bm r})$ maps at 5 different $I$ levels (H1-H5) with the same color scale as in Fig. 3.  (d) The $T(x)$ length profiles  corresponding to H1-H5 shown in Fig. 3(c).}
\label{fig.5}
\end{figure*}

\subsection{Non-mutualistic hot-spot and radiation coexistence}
For the same mesa A at $T_{\rm bath}$ = 55 K, the IVCs and the radiation characteristics are presented in Figs. 5(a) and 5(b), and the associated $T({\bm r})$ maps observed at bias points H1-H5 are shown in Fig. 5(c).  Although  panels H1 and H2 show some weak hot-spot $T({\bm r})>T_c$ regions, panels H3-H5 do not.  The $T(x)$ length profiles for panels H1-H5 shown in Fig. 5(d) are more uniform at $T_{\rm bath}$ = 55 K than are the same mesa profiles at the lower $T_{\rm bath}$ = 35 K shown in Fig. 3(e).  Nonetheless, Fig. 5(b) shows that the sub-THz radiation observed by the Si-bolometer near the retrapping points corresponding to panels H4 and H5, is as strong as that observed in the high-$I$ bias region at $T_{\rm bath}$ = 35 K.  Among the hundreds of mesas studied, the strongest emission of a few tens of $\mu$W was observed in such low-$I$ bias retrapping regions at around $T_{\rm bath}$ = 55 K\cite{Sekimoto,Yamamoto}.  The sub-THz frequency measurements show that $f=f_J=f_{\rm cav}$ either in the high-$I$ bias hot-spot region at low $T_{\rm bath}$ values or in the low-$I$ bias retrapping region at high $T_{\rm bath}$ values.  These observations clearly support the hypothesis that a hot spot neither enhances the radiation intensity nor alters the relation $f=f_J=f_{\rm cav}$ for the intense emission.  In other words the hot spot and the emission exhibit a non-mutualistic coexistence, contrary to previous interpretations\cite{Wang1,Wang2,Guenon,Li}.

Besides, in the high-$I$ bias regime, it was shown experimentally that the line width $\Delta f$ decreases dramatically with increasing $T_{\rm bath}$\cite{Li} and decreasing the dc input power $P_{\mathrm{dc}}$, implying that the IJJ synchronization increases dramatically with increasing $T_{\rm bath}$ and decreasing $P_{\mathrm{dc}}$.   If hot spots were the main source of the  remarkable high-$I$ bias synchronization,  one would expect that stronger hot spots would lead to greater synchronization.  Thus, we conclude that the hot spot probably does not provide the primary mechanism for the synchronization of the IJJs for the sub-THz emission.   Although previous workers presented convincing evidence that the overall emission line width was by far the narrowest in the high-$I$ bias R-type regions such as pictured in panels H5 of Fig. 3(d) containing a hot spot, and concluded that the hot spot might account for the very high degree of synchronization of the emission from the many IJJs\cite{Li}, our experiments and the simulations of Gross {\it et al.} do not provide support for that conclusion\cite{Gross}.  It seems reasonable to suppose that the normal central hot-spot region in the mesa in the high-$I$ bias region would decrease the overall coherence of the radiation emitted from the two separate superconducting regions, unfavorable for the synchronization between the two divided superconducting regions.  However, we still can't completely exclude the possibility that a hot spot might alter the sub-THz emission synchronization conditions, such as the impedance matching conditions, since (a) synchronization processes are highly non-linear and thus poorly understood, and (b) the hot spot could act as a variable shunt resistor, modifying the effective impedance for the radiation\cite{Yurgens}, \textit{etc}.

\subsection{Non-observation of $T({\bm r})$ standing waves}
More specifically, our SiC microcrystalline PL measurement technique did not provide any signs of standing-wave formation in $T({\bm r})$ maps between $T_{\rm bath}$ =15 K and 60 K\cite{Watanabe}, although only two of those map sets are shown here.  These results are in sharp contrast to those obtained by LTSLM and in some high-$I$ embedded-film observations, in which those different experimental techniques also observed standing waves in addition to hot spots\cite{Wang1,Wang2,Guenon,Benseman13}.  Gu\'enon \textit{et al.} suggested that the wavy features in the LTSLM images could be regarded as standing magnetic field waves\cite{Guenon}.  Wang \textit{et al.} suggested that an edge of the hot spot could act as a reflective cavity termination plane responsible for such standing waves forming along the length of the rectangular mesa\cite{Wang2}.  This propagation direction is inconsistent with our observations, because the radiation frequency $f$ nearly always obeys the TM(1,0) EM cavity resonance frequency $f_{\rm cav}=c_0/(2nw)$ for a standing wave across the rectangular mesa width $w$\cite{Kashiwagi}, independent of the size and location of the hot spot.  Furthermore, since the length of the superconducting region of the mesa, the effective EM cavity dimension $\ell_{\rm eff}$ in their model, becomes larger as $I$ is decreased, $f$ should decrease continuously, in contrast with their observations\cite{Wang2}.  In our observations $f$ remains invariant even during a sudden jump in the hot-spot center position\cite{Watanabe}.  In our simultaneous spectroscopic radiation observations, $f$ is completely unrelated to the hot spot's size and position, but is primarily determined  by the $ac$-Josephson relation $f=f_J$ and secondarily by $f=f_{\rm cav}$.  In the present case the emission was observed for 0.41 $\le f\le$ 0.76 THz.

\subsection{EM cavity resonance enhancements}
Finally, we discuss briefly the secondary condition important for the intensity of the sub-THz radiation: the rectangular EM cavity resonance condition $f=f_{\rm cav}=c_0/(2nw)$ for the rectangular mesa.  Hence, $f_{\mathrm{cav}}$ for mesas of the same $n$ is thus determined purely by $w$.  However, since the rather poor quality $Q$-value of the EM cavity is of the order of $\sim$10 due to the trapezoidal cross section of the mesa width, the actual EM cavity resonance has a broad frequency characteristic, approximately $\sim$50 GHz at the central frequency of 0.5 THz\cite{Kashiwagi}.  In the present mesa $w$ varies from 79 $\mu$m at the top of the mesa to 89 $\mu$m at the bottom, leading to the expected range of $f_{\mathrm{cav}}$ values from 0.42 to 0.48 THz.  In fact as seen by comparing Figs. 4(b) and 4(c), this secondary radiation mechanism appears to work quite well, since the strongest enhancement of the radiation is observed in this limited $f$ region, which is shown by the shaded yellow highlighted region in Fig. 4(c).

However, in contrast to this scenario, much weaker radiation can be observed both above 55 mA and below 22 mA.  For these faint radiation regions, the cavity resonance condition appears to be violated, even though the ac-Josephson relation is still fulfilled.  The THz radiation occurring without a good cavity resonance condition can naturally be understood by a mismatch of the standing wave of length $2w$ across the mesa width on which the above formula for $f_{\mathrm{cav}}$ was based.  This confirms that the cavity resonance condition only plays a secondary role in the THz radiation mechanism, as shown previously\cite{Tsujimoto1,Tsujimoto2,Delfanazari}.  Note that it also works as an important component to the overall mechanism for the synchronization of the IJJs and the power enhancement.

\section{summary}
In conclusion, an extremely inhomogeneous temperature distribution of the sub-THz emitting mesa consisting of the intrinsic Josephson junctions in the high-$T_c$ superconductor $\mathrm{Bi_2Sr_2CaCu_2O_{8+\delta}}$ was directly observed from the high spatial resolution of the  photoluminescence of many uniformly attached SiC microcrystals during simultaneous spectroscopic sub-THz radiation measurements.  This hot-spot temperature distribution was free of any apparent standing waves. Since the  hot spots  weaken the sub-THz radiation power, the two phenomena appear to coexist non-mutualistically.  A hot-spot phase synchronization mechanism therefore appears improbable.

\begin{acknowledgments}
 The authors thank Mitsuhiro Muroi for his assistance in data analysis and H. Asai, R. Kleiner,  W.-K. Kwok, H. B. Wang, and U. Welp  for fruitful discussions. This work was  supported in part  by the Grant-in-Aid for Challenging Exploratory Research from the Ministry of Education, Culture, Sports, Science and Technology.

\end{acknowledgments}


\begin{thebibliography}{99}
\bibitem{cavitymode} In general, the electromagnetic cavity excitation modes can be written as $f_{\mathrm{cav}}=c_0 k'_{\mathrm{mp}}/2\pi n_{\mathrm{r}}$, where $k'_{\mathrm{mp}}=\pi\sqrt{(m/w)^2+(p/L)^2}$, assuming that the rectangle has short and long edges with lengths $w$ and $L$, respectively, $m$ and $p$ are positive integers, $c_0$ is the light velocity in vaccum, $n_{\mathrm{r}}$ is the refractive index inside the cavity ($\approx$4.2 in the case of Bi-2212).  The mode is uniform in the thickness direction because it is very thin in comparison with the other dimensions.
\end{thebibliography}

\end{document}